\documentclass[12pt,dvips]{article}
\usepackage{amsmath,amssymb}
\usepackage{jcp}
\def\baselinestretch{1.00}
\usepackage{epsfig}
\begin{document}
\title{Implementation of a Quantum Search \\ Algorithm
on a Nuclear Magnetic Resonance \\ Quantum Computer}
\author{}
\maketitle
\begin{quotation}
\raggedright\noindent\frenchspacing

Jonathan A. Jones*\dag, Michele Mosca\dag\ddag, Rasmus H. Hansen\dag\\

* Oxford Centre for Molecular Sciences, New Chemistry Laboratory,
South Parks Road, Oxford, OX1~3QT, UK,\\
\dag\ Centre for Quantum Computation, Clarendon Laboratory, Parks Road,
Oxford OX1~3PU, UK\\
\ddag\ Mathematical Institute, 24--29 St Giles', Oxford, OX1~3LB, UK\\
\end{quotation}

\textbf{
The simulation of quantum mechanical systems with classical computers 
appears to be a
computationally intractable problem.  In 1982 Feynman\cite{Feynman82}
reversed this observation, suggesting that quantum mechanical systems have
an information processing capability much greater than that of
corresponding classical systems, and thus could be used to implement a new
type of powerful computer.  In 1985 Deutsch\cite{Deutsch85} described a
quantum mechanical Turing machine, showing that quantum computers could
indeed be constructed.  Since then there has been extensive research in
this field, but while the theory is fairly well understood actually building a
quantum computer has proved extremely difficult, and only two methods have
been used to demonstrate quantum logic gates: ion
traps\cite{Cirac95,Monroe95}, and nuclear magnetic resonance
(NMR)\cite{Cory97,Chuang97}.  NMR quantum computers have previously been
used to demonstrate quantum algorithms to solve the two bit Deutsch
problem\cite{Jones98,Chuang98}.  Here we show how such a computer can be
used to implement a fast quantum search algorithm initially developed by
Grover\cite{Grover96,Grover97}.
}

Among other applications Grover's algorithms enable
an extremely rapid search over the domain of a binary function to find
elements for which this function is satisfied (that is, the function has
the value $1$).  This approach is simpler if the number of satisfying
values is known beforehand, and is particularly simple when precisely
one quarter of the elements in the domain satisfy the function\cite{BBHT96}.
The algorithm can be demonstrated using a computer with two quantum bits
(qubits) to search a two bit domain in which one of the four elements
satisfies the function.  A classical search of this domain would require
between 1 and 3 evaluations of the function to find the satisfying element,
while a quantum search can find this element with only one function
evaluation.  In the more general case of searching a domain of size
$N$ for one of $k$ satisfying elements the classical search requires
about $\frac{1}{2}(N/k)$ function evaluations,
while the quantum search\cite{BBHT96} requires only $O(\sqrt{N/k})$.

A quantum circuit for implementing a quantum search in a 
two qubit system is shown in Fig.~\ref{fig:algorithm}.  This
algorithm uses a single function evaluation to label the single
state which satisfies the function, followed by a series of gates
which drive the system into this particular state.
There are four possible functions $f$ with a unique satisfying element,
and these functions are conveniently labelled by the bit pattern
of the satisfying element.  The algorithm starts with the quantum computer
in state $|00\rangle$ and ends with the computer in a state corresponding
to the bit pattern of the unique element, and so this element can be
immediately identified by determining the final state of each qubit.

This algorithm was implemented using our
two qubit NMR quantum computer, described in Ref.~\citen{Jones98}.  This
computer uses the two spin states of ${}^1\rm H$ nuclei in a magnetic
field as qubits, while radio frequency (RF) fields and spin--spin
couplings between the nuclei are used to implement quantum logic gates.
Our molecule, partially deuterated cytosine, contains two  ${}^1\rm H$
nuclei, and thus can be used to implement a two qubit computer.
The pseudo-Hadamard gates ($h$) were implemented using $90^\circ_y$ pulses,
while the function evaluation was performed using the pulse sequence
shown in Fig.~\ref{fig:Ufab}.
The phases of five of the pulses depend on which of the four possible
functions is to be implemented, and these phases should be set as shown in
the figure caption.  The final gate, $U_{00}$, is easily implemented, as it
is identical to $U_{f_{00}}$.

The algorithm should start with the computer in state $|00\rangle$, but
with an NMR quantum computer it is not practical to begin in a true
$|00\rangle$ state.  Using the methods of Cory {\it et al.}\cite{Cory97}
it is, however, possible to create an effective pure
state, which behaves in an equivalent manner.  Similarly it is not practical
to determine the final state directly, but an equivalent measurement
can be made by exciting the spin system with a further $90^\circ$ pulse,
and observing the phases of the resulting NMR signals.  The absolute phase
of an NMR signal depends in a complex manner on a variety of experimental
details, and so it is not possible to interpret absolute phases, but this
can be overcome by measuring a reference signal, obtained by applying
a $90^\circ$ pulse directly to the initial state.

\begin{figure}
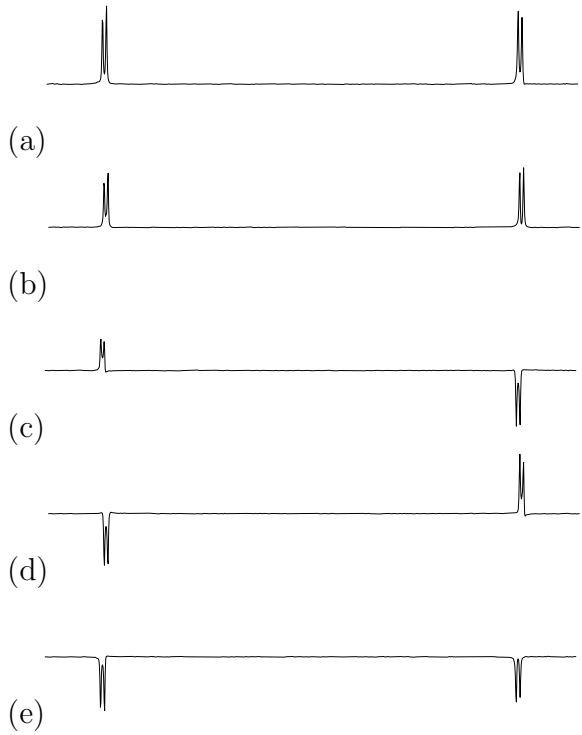

\mbox{}\\
(a)\epsfig{file=ref.ps}\\
(b)\epsfig{file=f00.ps}\\
(c)\epsfig{file=f01.ps}\\
(d)\epsfig{file=f10.ps}\\
(e)\epsfig{file=f11.ps}\\
\caption{Experimental spectra from our NMR quantum computer.  Spectrum (a)
is a reference spectrum, used to determine the absolute phases of the NMR
signals, while spectra (b)--(e) were acquired from the same computer
implementing the quantum search algorithm using each of the four possible
functions: (b) $f_{00}$, (c) $f_{01}$, (d) $f_{10}$, (e) $f_{11}$.  These
four spectra were processed using the reference phase obtained from spectrum
(a), and so the phases of the signals can be interpreted as states of the
corresponding qubits, as described in the main text.
\label{fig:spectra}}
\end{figure}

The results of this approach are shown
in Fig.~\ref{fig:spectra}.  Five spectra are shown: a reference spectrum
acquired using a single $90^\circ$ pulse, and spectra acquired
from the same computer implementing the search algorithm for each of the
four possible functions, $f$.  Each spectrum consists of two closely
spaced pairs of lines: each pair of lines corresponds to a single qubit,
while the barely visible splitting within each pair arises from the
spin--spin coupling, $J$, used to implement the two qubit gates.
To improve the appearance of the spectra the final $90^\circ$ detection
pulse was preceded by a magnetic field gradient pulse, which acts to
dephase the majority of any error terms which might occur.

The reference spectrum corresponds to the computer being in state
$|00\rangle$, and the phase of this spectrum was adjusted so that both lines
are in positive absorption phase (that is, pointing upwards).  The same
phase correction was then applied to the other four spectra, allowing
positive absorption lines to be interpreted as qubits in state
$|0\rangle$, while negative absorption lines can be interpreted as qubits
in state $|1\rangle$.  The left hand pair of lines arises from the first
spin, and thus corresponds to the first qubit, while the right hand pair
of lines corresponds to the second qubit.  Thus, for example, spectrum (c)
corresponds to the state $|01\rangle$.
Examining the four spectra (b)--(e), it is clear that our implementation
of a quantum search using the function $f_{ab}$ leaves the
computer largely in a final state $|ab\rangle$, much as expected.

There are however small but significant errors in the calculation, which
result in distortions in the final spectra.  While most of these distortions
are removed by the field gradient pulse, their effects remain visible as
variations in the heights of the NMR lines.
These errors arise from a variety of causes, but the most significant is
errors in the NMR pulse sequence used to implement the calculation.  These
pulse sequences require a large number of selective pulses, that is pulses
which only affect one of the two spins making up the computer.  In practice
it is difficult to achieve the desired effect at one spin while leaving
the other entirely unaffected, resulting in errors in the final result.
Interestingly the distortions are much worse in some cases than in others:
they are particularly bad in the spectrum obtained when the function is
$f_{01}$.  Understanding this variability may lead to techniques for reducing
the distortion in all cases.

Implementing Grover's algorithm is a major step forward for NMR quantum
computing (since this letter was first submitted an implementation of
Grovers's algorithm using heteronuclear NMR has been
published\cite{Chuang98b}), but is by no means the limit of what can be
achieved.  Preliminary studies have been made of systems containing
three qubits\cite{Laflamme98}, and it should be possible to build larger
NMR quantum computers, allowing the implementation of more complex algorithms.

\noindent{\sc acknowledgments.}
We thank A. Ekert for helpful discussions.  J.A.J. thanks C. M. Dobson
for his encouragement.  This is a contribution from the Oxford Centre for
Molecular Sciences which is supported by the UK EPSRC, BBSRC and MRC.
M.M. thanks CESG (UK) for their support.  R.H.H. thanks the
Danish Research Academy for financial assistance.

\noindent{\sc correspondence} should be addressed to J.A.J. (e-mail:
jones@bioch.ox.ac.uk).

\begin{figure}
\begin{picture}(235,80)
\put(10,60){\makebox(0,0)[r]{$|0\rangle$}}
\put(15,60){\line(1,0){10}}
\put(25,50){\framebox(20,20){$h^{-1}$}}
\put(45,60){\line(1,0){15}}
\put(10,20){\makebox(0,0)[r]{$|0\rangle$}}
\put(15,20){\line(1,0){10}}
\put(25,10){\framebox(20,20){$h^{-1}$}}
\put(45,20){\line(1,0){15}}
\put(60,10){\framebox(30,60){$U_{f_{ab}}$}}
\put(90,60){\line(1,0){15}}
\put(105,50){\framebox(20,20){$h$}}
\put(125,60){\line(1,0){15}}
\put(90,20){\line(1,0){15}}
\put(105,10){\framebox(20,20){$h$}}
\put(125,20){\line(1,0){15}}
\put(140,10){\framebox(30,60){$U_{00}$}}
\put(170,60){\line(1,0){15}}
\put(185,50){\framebox(20,20){$h^{-1}$}}
\put(205,60){\line(1,0){10}}
\put(220,60){\makebox(0,0)[l]{$|a\rangle$}}
\put(170,20){\line(1,0){15}}
\put(185,10){\framebox(20,20){$h^{-1}$}}
\put(205,20){\line(1,0){10}}
\put(220,20){\makebox(0,0)[l]{$|b\rangle$}}
\end{picture}
\caption{A quantum circuit for the implementation of a quantum search
algorithm on a two qubit computer.  Gates marked $h$ (pseudo-Hadamard
gates) act to take a single
eigenstate to a uniform superposition of the four possible eignestates,
while gates marked $h^{-1}$ implement the inverse operation.  The first
two qubit gate $U_{f_{ab}}$ corresponds to an evaluation of the function
$f_{ab}$, replacing an eigenstate $|ij\rangle$ by $-|ij\rangle$ if
$i=a$ and $j=b$, while $U_{00}$ simply replaces $|00\rangle$ by $-|00\rangle$.
\label{fig:algorithm}}
\end{figure}
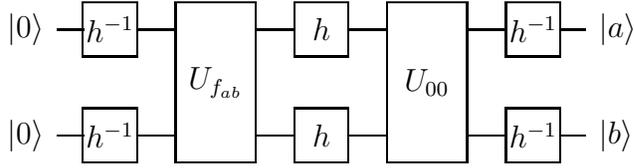
\begin{figure}
\epsfig{file=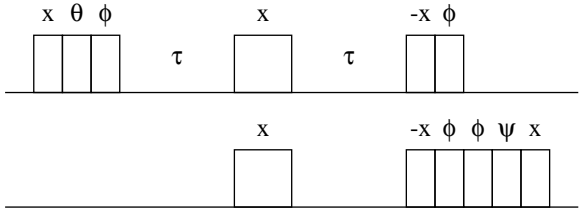}
\caption{NMR pulse sequence used to implement $U_{f_{ab}}$.  Narrow boxes
correspond to $90^\circ$ pulses, while wide boxes are $180^\circ$ pulses;
the upper and lower lines refer to the nuclear spins corresponding to the
first and second qubits respectively.  The time period $\tau$, during which
no pulses are applied, is set equal to $1/4J$, where $J$ is the size of
the spin--spin coupling between the nuclei.  The phase of each pulse is
written above it, and for five of the pulses this phase depends on 
which of the four functions $f_{ab}$ is to be implemented.  These phases
should be set as follows:
$f_{00}$, $\theta=+y$, $\phi=+x$, $\psi=-y$;
$f_{01}$, $\theta=+y$, $\phi=-x$, $\psi=+y$;
$f_{10}$, $\theta=-y$, $\phi=-x$, $\psi=-y$;
$f_{11}$, $\theta=-y$, $\phi=+x$, $\psi=+y$.
\label{fig:Ufab}}
\end{figure}

\end{document}